\documentclass[aps,prd,superscriptaddress,twocolumn,
amsmath,amssymb,nofootinbib]{revtex4-1}

\usepackage{slashed}
\usepackage{graphicx}
\usepackage{color}
\usepackage[breaklinks,colorlinks=true]{hyperref}
\usepackage{cancel}
\usepackage[normalem]{ulem}

\newcommand{\bea}{\begin{eqnarray}}
\newcommand{\eea}{\end{eqnarray}}
\newcommand{\beq}{\begin{equation}}
\newcommand{\eeq}{\end{equation}}

\newcommand{\bega}{\begin{eqnarray}}
\newcommand{\eega}{\end{eqnarray}}

\newcommand{\sn}{\mathtt{sn}}
\newcommand{\cn}{\mathtt{cn}}
\newcommand{\dn}{\mathtt{dn}}
\newcommand{\am}{\mathtt{am}}

\DeclareMathSymbol{\Epsilon}{\mathalpha}{operators}{"45}
\DeclareMathSymbol{\Kappa}{\mathalpha}{operators}{"4B}

\begin{document}
\title{Quantum vacuum, rotation, and nonlinear fields}

\author{Antonino Flachi}
\affiliation{Department of Physics,
  \& Research and Education Center for Natural Sciences,
  Keio University, 4-1-1 Hiyoshi, Yokohama, Kanagawa 223-8521, Japan}
\email{flachi@phys-h.keio.ac.jp}
  
\author{Matthew Edmonds}
\affiliation{ARC Centre of Excellence in Future Low-Energy Electronics Technologies,
School of Mathematics and Physics, University of Queensland, St Lucia, QLD 4072, Australia}
\email{m.edmonds@uq.edu.au}
\affiliation{Department of Physics,
  \& Research and Education Center for Natural Sciences,
  Keio University, 4-1-1 Hiyoshi, Yokohama, Kanagawa 223-8521, Japan}

\begin{abstract}
In this paper, we extend previous results on the quantum vacuum or Casimir energy, for a noninteracting rotating system and for an interacting nonrotating system, to the case where both rotation and interactions are present. Concretely, we first  reconsider the noninteracting rotating case of a scalar field theory and propose an alternative and simpler method to compute the Casimir energy based on a replica trick and the Coleman-Weinberg effective potential. We then consider the simultaneous effect of rotation and interactions, including an explicit breaking of rotational symmetry. To study this problem, we develop a numerical implementation of zeta function regularization. Our work recovers previous results as limiting cases and shows that the simultaneous inclusion of rotation and interactions produces nontrivial changes in the quantum vacuum energy. Besides expected changes (where, as the size of the ring increases for fixed interaction strength, the angular momentum grows with the angular velocity), we notice that the way rotation combines with the coupling constant amplifies the intensity of the interaction strength. Interestingly, we also observe a departure from the typical massless behavior where the Casimir energy is proportional to the inverse size of the ring. 
\end{abstract}
\maketitle

\section{Introduction}
Consider a periodic array of identical particles at low temperature confined to a one-dimensional ring by means of {a particular} trapping potential. Then, making such a system rotate at a constant velocity, i.e. uniformly, is a trivial exercise. What we mean by ``trivial'' here is that an observer at rest in the laboratory frame cannot detect that the device is rotating: the system is invariant under a rotation of reference frames. Put in a more elegant way, the gauge invariance associated with rotation is preserved. Obviously, if this gauge invariance is in some way broken, then the above is no longer true and rotation acquires a physical, in principle measurable, element or, in other words, the observer at rest in the laboratory frame can detect rotation. It is nontrivial, in fact interesting, to build up models where this symmetry can be broken; however, in $1+1$ dimensions the Coleman-Hohenberg-Mermin-Wagner theorem prevents this from happening dynamically \cite{Mermin:1966,Hohenberg:1968,Coleman:1973}. The simplest and most natural way to break this gauge invariance is to do this \textit{classically} by enforcing nontrivial boundary conditions at one point along the ring, thus breaking the periodicity of the array: this is equivalent to an explicit symmetry breaking. A simple way to achieve this is by placing an impurity somewhere along the ring, thus breaking the uniformity of the array. A pictorial explanation of this is given in Fig.1. 

The ``device'' we have described above, whether rotating or static, is susceptible {to} a Casimir force, arising from the deformations of its (quantum) vacuum due to the compactness of the topology of the setup, i.e., the boundary conditions \cite{Milton:2001,Bordag:2009}. If the ring is static ({i.e.,} non rotating) and periodic ({i.e.,} periodic boundary conditions are imposed at \textit{any} one point along the ring), then quantum fluctuations are massless ({i.e.,} long-range) and induce a Casimir force that scales with the inverse size of the ring. Then, the arguments given in the previous paragraph imply that making such a periodic ring rotate should not change the Casimir force. However, if we change the boundary conditions into nonperiodic ones, something interesting happens: rotation becomes ``physical'', quantum fluctuations have to obey nontrivial boundary conditions, and as a result the Casimir force should respond to variations in angular velocity.

A computation of the Casimir energy for a noninteracting scalar field theory can be found in Refs.~\cite{Chernodub:2012em,Schaden:2012}, while a computation of the Casimir energy for a scalar field theory with quartic interactions can be found in Ref.~\cite{Bordag:2021}. A calculation of the one-loop effective action for a non-relativstic nonlinear sigma model with rotation can be found in Ref.~\cite{Corradini:2021yha}. Here, we consider the simultaneous effect of both rotation and interactions.

The paper is organized as follows. After introducing the notation by illustrating the calculation of the Casimir energy for a real scalar field, we proceed by reviewing the free-field rotating case and illustrate an unconventional and computationally convenient way to calculate the Casimir energy. We then move on to the interacting, rotating case and discuss how both factors alter rather nontrivially the Casimir energy.

\begin{figure}
\begin{center}
\includegraphics[scale=0.25,clip]{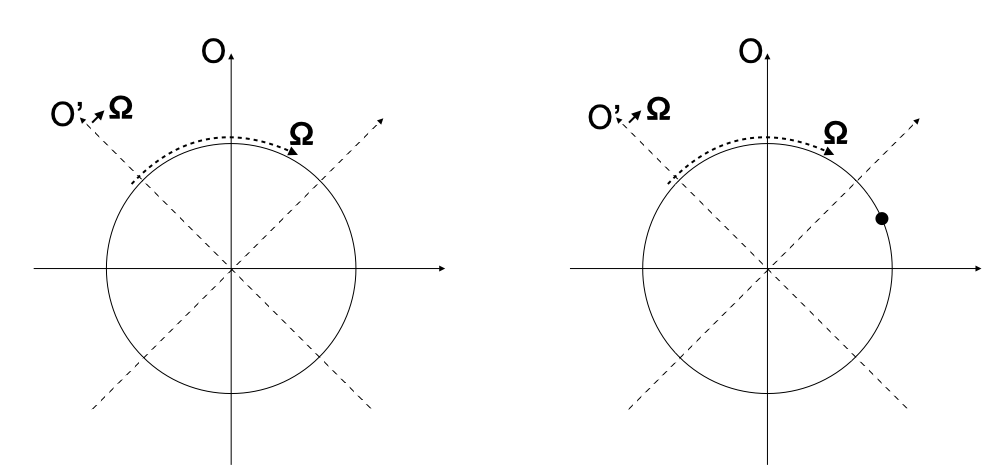}
\end{center}
\caption{Both panels represent a ring rotating with an angular velocity $\Omega$. On the left, periodic boundary conditions are imposed at any one point, and the two observers (represented by the two reference frames $O$ and $O'$, the latter rotating with the ring) cannot detect that the ring is rotating. The ring on the right has an impurity (represented by a black dot) that breaks the rotational symmetry. In this second case, for the observer $O'$ at rest with the ring (i.e., rotating with the ring) the position of the impurity would not change (i.e. the boundary conditions are time independent), while for the \textit{laboratory} observer $O$, rotation is evident and the boundary conditions imposed at the impurity are time dependent.}
\label{figure}
\end{figure}

\section{Free case}

To introduce our notation, let us consider the case of a noninteracting {single}-component, real scalar field, whose Lagrangian density, in absence of rotation, is
\bea
\mathcal{L} 
   =  {1\over 2} \left[ \left({\partial \phi\over \partial \hat t}\right)^2 - {1\over R^2}\left({\partial \phi\over \partial \hat \varphi}\right)^2\right].
\eea
$R$ is the radius of the ring and $\hat \varphi \in \left[0,2\pi\right)$. In the first part of this section we simply repeat the calculation of Refs.~\cite{Chernodub:2012em,Schaden:2012}; in the second part of this section, we carry out the calculation of the vacuum energy in a simpler way. The field $\phi$ is assumed to satisfy certain boundary conditions at the end points, $\hat\varphi=0,~2\pi$. Taking for concreteness Dirichlet  boundary conditions, we have
\bea
\phi(\hat t,\hat \varphi)\Big|_{\hat \varphi=0}
=
\phi(\hat t,\hat \varphi)\Big|_{\varphi=2\pi}
=0.
\label{DirichletBC}
\eea
In this case the system is static and a potential barrier at $\hat \varphi=0=2\pi$ enforces the above Dirichlet boundary conditions. Changing the potential barrier will change the boundary conditions. The Casimir energy is defined as 
\bea
E = \int_{0}^{2\pi R} dx \rho(x),
\eea
where $\rho(x) = \langle T_{00} (x)\rangle$ is the energy density component of the energy-momentum tensor in the vacuum state. 
The expression above can be written as the \textit{regularized} sum over the zero-point energy levels $E_n$ of the fluctuations,
\bea
E = \sum_{n} {}^{'} E_n,
\eea
which defines the Casimir energy. The prime in the sum is a reminder that the summation is divergent and requires regularization. Throughout our paper we adopt zeta function regularization \cite{Toms:2012}. A straightforward calculation of the Casimir energy gives \cite{Bordag:2009}
\bea
E = -\pi/(24 \times L),
\label{Ef}
\eea
showing that the energy scales as the inverse size of the ring $L =2 \pi R$. This is the one-dimensional scalar equivalent of the usual electromagnetic Casimir energy; see e.g. Ref.~\cite{Milton:2001}. The numerical coefficient and the sign in Eq.~(\ref{Ef}) depend on the boundary conditions, on the features of the vacuum fluctuations (e.g., spin) and of the background (e.g., an external potential or a nontrivial background geometry in more than one dimension). In the present case the Casimir force is attractive. 

\section{Spinning the ring}

In this section we describe what happens to the Casimir energy when the ring is rotating. In the first part of this section we review some known results, Refs.~\cite{Chernodub:2012em,Schaden:2012}, and compute the Casimir energy using the standard way of summing over the eigenvalues and regularizing the sum. In the second part of the section we show how it is possible to obtain the same result for the Casimir energy {in a different way} using a replica trick applied to functional determinants. 

Following the standard approach, we first pass from the laboratory frame (where the boundary conditions are time dependent) to a corotating frame (where the boundary conditions are simply Eq.~(\ref{DirichletBC})) by performing the following coordinate transformation,
\bea
t = \hat {t},~~~ \varphi =\hat{\varphi} + \Omega t.
\eea
\bea
{\partial \over \partial t} &\to& {\partial \over \partial \hat t}  - \Omega {\partial \over \partial \hat\varphi}, \nonumber\\
{\partial \over \partial \varphi} &\to& {\partial \over \partial \hat \varphi}, \nonumber
\eea
that leads to the following Lagrangian density:
\bea
{L} = 
 {1\over 2} \left[ 
 \left(
 {\partial  \phi\over \partial t} +\Omega {\partial  \phi \over \partial \varphi}
 \right)^2 
 - {1\over R^2}\left({\partial \phi\over \partial  \varphi} \right)^2
 \right].
 \label{lagr}
\eea
The equation of motion following from the above Lagrangian density are 
\bea
0 = \left(
\left({\partial \over \partial t} -\Omega {\partial  \over \partial \varphi}\right)^2
-{1\over R^2}{\partial^2 \over \partial \varphi^2} 
\right) \phi .
\label{eq6}
\eea
Imposing Dirichlet boundary conditions means
\bea
\phi \left(t, \varphi\right) \Big|_{\varphi = 0, 2\pi}= 0.
\label{eq7}
\eea
Notice that the above boundary conditions are time independent for a corotating observer. 

First, {we describe} the ``canonical'' method. 
Assuming that any solution to Eq.~(\ref{eq6}) can be written as
\bea
\psi(t, \varphi) = \sum_n a_n e^{-i \omega_n t} f_n(\varphi) + a^\dagger_n e^{i \omega_n t} f^*_n(\varphi).
\label{dec_norm_mods}
\eea
with $\omega_n \geq 0$,
and substituting the above expression into the original equation for $\psi$, we obtain
\bea
0 &=& 
\sum_n \omega_n^2 e^{-i \omega_n t} f_n(\varphi)
- \left(\Omega^2-R^{-2}\right) \sum_n e^{-i \omega_n t} f''_n(\varphi) \nonumber \\
&&- 2 i \Omega \sum_n \omega_n e^{-i \omega_n t} f'_n(\varphi)
\eea
The above assumption is verified if the modes $f_n(\varphi)$ are a complete and orthogonal set of solutions satisfying {the} equation
 \bea\label{eqn:fn}
0 = 
\left(1 - \beta^2\right)  f''_n(\varphi)
- 2 i \beta  \tilde\omega_n f'_n(\varphi)
+\tilde\omega_n^2  f_n(\varphi).
\label{eq12}
\eea
We have defined, for brevity of notation,
\bea
\beta^2 &=& \Omega^2 R^2,\nonumber\\
\tilde\omega_n &=& R\omega_n.
\eea
Equation \eqref{eqn:fn} can be solved exactly and after imposing the boundary conditions (\ref{eq7}) on the general solutions, one finds through simple algebra that the following quantization condition holds
\bea
\sin^2\left(
 {2 \pi \tilde \omega_n\over 1 - \beta^2} 
\right) = 0,
\eea
which gives the following spectrum of the quantum fluctuations
\bea
\omega_n ={n\over 2R }\left(1 - \beta^2\right),
\label{omega0n}
\eea
with $n \in \mathbb{N}$. The corresponding eigenfunctions can be written as follows
\bea
\phi_n(t,\varphi) =   {1\over \sqrt{\pi R}} e^{-i \omega_n t} e^{\imath \varphi {\beta \omega_n R \over 1- \beta^2}}  \sin\left({n \varphi \over 2}\right),
\eea
with the prefactor coming from the requirement that the modes are normalized.
The above solutions for the modes $\phi_n(t, \varphi)$ correspond to the modes in the corotating frame. To go back to the stationary-laboratory frame, we can perform the inverse coordinate transformation,
\bea
t &\to& t, ~~~~~~~~~\mbox{and}~~~~~~~~~ \varphi \to \varphi + \Omega t \nonumber
\eea
to get
\bea
\phi_n(t,\varphi) =   {1\over \sqrt{\pi R}} e^{-i \omega_n t} e^{\imath \left[\varphi + \Omega t\right]_{2\pi} {\beta \omega_n R \over 1- \beta^2}}  \sin\left({n \over 2} \left[\varphi + \Omega t\right]_{2\pi}\right),\nonumber
\eea
with $\left[ u \right]_{2\pi} \equiv u \left[\mbox{mod($2\pi$)}\right]$, thus explicitly indicating that the solutions have a $2\pi$ periodicity. These solutions and the method we have described are those of Refs.~\cite{Chernodub:2012em,Schaden:2012}.

A direct way to compute the Casimir energy is to perform the regularized sum over the eigenvalues. It is straightforward to write
\bea
E_r &=& {1\over 2} \lim_{s \to -1}\sum_{n=1}^\infty \left({n\over 2R }\left(1 - \beta^2\right)\right)^{-s} \\
&=& -{1\over 2\times 12\times 2 R }\left(1 - \beta^2\right),
\eea
where the factor $1/12$ comes from the summation over $n$ yielding $\zeta(-1)=-1/12$. Here, $\zeta(s)$ defines the Riemann zeta function.

As clearly discussed in Refs.~\cite{Chernodub:2012em,Schaden:2012}, the Casimir energy $E_s$ in the stationary-laboratory frame is related to the Casimir energy $E_r$ in the rotating frame by the following relation, 
\bea
E_s - E_r =  {\bf \Omega L},
\eea
from which, by taking the inverse Legendre transform of $E_r$, one can obtain the angular momentum dependence of the ground state energy:
\bea\label{eqn:am}
L &=& -{\partial E_r \over \partial \Omega} 
=-{1\over 24} \Omega R.
\eea
It follows that
\bea
E_s=  -{1\over 48 R}\left(1 + \beta^2\right).
\label{Esd}
\eea
As noted in Ref.~\cite{Schaden:2012}, the above quantity is the quantum vacuum energy and it should be augmented by a classical contribution proportional to the moment of inertia, $I$, yielding the total energy to be
\bea
E_s=  -{1\over 48 R}\left(1 + \beta^2\right) +{1\over 2} I \Omega^2.
\eea
While Ref.~\cite{Chernodub:2012em} had initially and correctly argued that the vacuum energy lowers the moment of inertia of such a system, Ref.~\cite{Schaden:2012} later argued that this never happens, at least within the range of validity of a semiclassical treatment.

\section{Functional determinants and the replica trick}

Before including interactions, we will illustrate an alternative way to compute the Casimir energy for the free case with almost no calculation by using a replica trick to obtain directly the energy $E_s$ in the laboratory frame without passing through the solution of the mode equation. This is interesting for two reasons. First, it is simple. Second, it offers at least in some cases a simple way to compute the quantum vacuum energy in stationary (although simple) backgrounds, which is usually complicated due to the mixing of space and time components in the metric tensor that yields nonlinear eigenvalue problems \cite{Fursaev,Fursaev:2011}.

The method we outline below is valid in setups more general than what we discuss here, assuming that there are no parity breaking interactions. The replica trick here stems from the fact that the Casimir energy of the rotating system must be an even function of the rotational velocity, i.e. it does not change if we invert the direction of rotation. This simple physical consideration allows us to greatly simplify the calculation. 

Rather {than} following the canonical approach of summing over the zero-point energies, we pass through the Coleman-Weinberg effective potential from which one can extract the Casimir energy.
The Coleman-Weinberg effective potential can be obtained starting from the following functional determinant
\bea
\delta \Gamma = \log \det \left(\left({\partial\over \partial t} -\Omega {\partial\over \partial \varphi}\right)^2 -{1\over R^2}{\partial^2\over \partial \varphi^2}\right).
\eea
The problem with the expression above is that the presence of first order derivatives in time makes the eigenvalue problem {nonlinear}~\cite{Fursaev}. Here we will bypass the nonlinearities using the following replica trick. First of all, we can express the determinant as
\bea
\delta \Gamma = \log A = \log \det \left(L_+ \times L_-\right)
\eea
having defined
\bea
L_\pm \left(\Omega\right) = {\partial\over \partial t} -\alpha_\pm {\partial\over \partial \varphi}
\eea
with 
\bea
\alpha_\pm = \Omega \pm {1\over R}
\eea
This gives
\bea\label{eqn:det}
A = \det \left(L_+ \left(\Omega\right) \right)\times \det \left(L_- \left(\Omega\right)\right).
\eea
The replica trick relies on the {assumption} that inverting the sense of rotation $\Omega \to -\Omega$, does not change the effective action. This is a \textit{physical} assumption, which is valid in our case. If parity breaking terms are present, one can modify the trick by adding up a residue. In our case, {we can express Eq.~\eqref{eqn:det} as}
\bea
A &=& 
\left(
\det \left(L_+ \left(\Omega\right) L_- \left(-\Omega\right) \right) \times
\left(L_- \left(\Omega\right) L_+ \left(-\Omega\right) \right)
\right)^{1/2} \nonumber\\
&=&
\det 
\left(
\left({\partial^2\over \partial t^2} - {1\over \rho_+^2} {\partial^2\over \partial \varphi^2}\right)
\left({\partial^2\over \partial t^2} - {1\over \rho_-^2} {\partial^2\over \partial \varphi^2}\right)
\right)^{1/2}
\label{prodea}
\eea
where
\bea
\rho_\pm^{-1} = \left|\Omega \pm {1\over R}\right|.
\eea
From the above relation we can factorize the effective action as follows:
\bea
\delta \Gamma &=&
{1\over 2} \log \det \left({\partial^2\over \partial t^2} - {1\over \rho_+^2} {\partial^2\over \partial \varphi^2}\right)\nonumber\\
&&+
{1\over 2} \log \det \left({\partial^2\over \partial t^2} - {1\over \rho_-^2} {\partial^2\over \partial \varphi^2}\right).
\label{replea}
\eea
Notice that we have commuted the operators $L_+$ and $L_-$. This practice requires some justification since a noncommutative or \textit{Wodzicki residue} may appear under some circumstances in the trace functionals of formal pseudodifferential operators. In the present case it makes no difference, but in more general situations care should be paid about this point \cite{Elizalde:1997nd,Dowker:1998tb,McKenzie-Smith:1998ocy}.

The replica trick has allowed us to express the effective action for the rotating system in terms of the effective action for an analogous system without rotation but with an effective radius,
\bea
\delta \Gamma_\pm &=&
{1\over 2} \log \det \left({\partial^2\over \partial t^2} 
- {1\over \rho_\pm^2} {\partial^2\over \partial \varphi^2}\right).
\eea
The above is nothing but the integral over the volume of the Coleman-Weinberg potential for a free theory; combining the $\delta \Gamma_\pm$ terms as in Eq.~(\ref{replea}) should return the Casimir energy after renormalization. {We outline the computation of the $\delta \Gamma_\pm$ terms for completeness}. Setting
\bea
x_\pm = \rho_\pm \varphi
\eea
we have
\bea
\delta \Gamma_\pm=
{1\over 2} \log \det \left({\partial^2\over \partial t^2} -  {\partial^2\over \partial x_\pm^2}\right),
\eea
with boundary conditions imposed at $x_\pm=0$ and $x_\pm=2\pi \rho_\pm$. For Dirichlet boundary conditions, the eigenvalues of the above differential operators are
\bea
E^{(\pm)}_n = \pi n/x_\pm.
\eea
Using zeta regularization, we get for the effective potential $V_\pm$ (i.e., the effective action divided by the volume)
\bea
V_\pm = {1\over 2 x_\pm} \sum_n E^{(\pm)}_n. 
\eea
{One can then show that}
\bea
V_\pm = -{1\over 96 \pi \rho_\pm^2}.
\eea
The contribution to the Casimir energy from the $+$ and $-$ terms summed up gives
\bea
E_s = 2 \pi R \left( V_+ + V_-\right) = -{1\over 2\times 24 R}{\left(1 + R^2 \Omega^2\right)}, 
\label{eq:cas0}
\eea
where the factor $1/2$ comes from the overall $1/2$ in the factorization of the determinant. The above formula coincides with Eq.~(\ref{Esd}), and leads to an attractive force. 
Adding additional flat directions, orthogonal to the rotating ring, turning it into a rotating cylinder, does not change the attractive nature of the force for the physically relevant regime of $\Omega R \ll 1$. If we change the boundary conditions to periodic, then the sums in the above zeta function $\zeta_\pm$ extend over $n \in \mathbb{Z}$, resulting in a total Casimir energy independent of the rotational velocity $\Omega$, as we argued earlier. 

\section{Adding rotation and interactions}
Now, we get to the physically novel part of this work, that is working out the vacuum energy in the presence of rotation and interactions. The nonrotating case has been worked out in Ref.~\cite{Bordag:2021} and here those results will be reobtained as a limiting case of our more general expressions. In the following, we shall consider a complex scalar theory with quartic interactions and start from the Lagrangian density
\bea
{L} = 
 {1\over 2} \left[ 
 \left(
 {\partial  \phi\over \partial t} +\Omega {\partial  \phi \over \partial \varphi}
 \right)^2 
 - {1\over R^2}\left({\partial \phi\over \partial  \varphi} \right)^2
+{\lambda \over 4} \phi^4 \right],~~~
 \label{lagr_int}
\eea
from which the equations of motion can be easily obtained,
\bea
0 = \left(
\left({\partial \over \partial t} -\Omega {\partial  \over \partial \varphi}\right)^2
-{1\over R^2}{\partial^2 \over \partial \varphi^2} 
+{\lambda} \phi^2
\right) \phi .
\label{eq_rot_int_0}
\eea

The computation of the Casimir energy in the present case can be carried out in a similar manner as we did in the free case (i.e., going over the mode decomposition, finding the spectrum of the fluctuations, and performing the renormalized sum over the zero-point energies); however, differently from before, in general the Casimir energy cannot be obtained in analytical form. The reason will become clear in a moment.

After proceeding with the decomposition in normal modes, as in Eq.~(\ref{dec_norm_mods}), we obtain the following nonlinear Sch\"odinger equation
\bea
0 = \left(
\left(-{\imath \omega_p} - \beta {\partial  \over \partial x}\right)^2
-{\partial^2 \over \partial x^2} 
+{\lambda} f_p^2
\right) f_p.
\label{eq_rot_int}
\eea
Above, we have defined $x = R \varphi$, and the index $p$ is a generic, boundary-conditions-dependent multi-index that can in principle take continuous and/or discrete values. Setting $\lambda=0$, which is in the noninteracting limit, returns Eq.~(\ref{eq12}).

Notice that Eq.~(\ref{eq_rot_int}) differs from the analogous equation of Ref.~\cite{Bordag:2021} by the addition of the term proportional to $\beta$; such a contribution combines with the frequency $\omega_p$ making the associated eigenvalue problem nonlinear, as we anticipated in the previous section.


For notational convenience we first write the mode equation as follows:
\bea\label{eqn:nlse}
0 = 
f_p'' - \imath a_p f_p' + b_p f_p - \gamma \left|f_p\right|^2 f_p,
\label{eq_damped_cubed_oscillator}
\eea
where the prime refers to differentiation with respect to $x$ and
\bea
a_p &=& {2  \omega \beta \over 1-\beta^2} > 0, \label{Eq:42}\\
b_p &=& {\omega^2\over 1-\beta^2} > 0, \label{Eq:43}\\
\gamma &=& {\lambda \over 1-\beta^2}. \label{Eq:44}
\eea
Equation \eqref{eqn:nlse} is that of a damped, cubed anharmonic oscillator and can be explicitly solved only for special values or combinations of the parameters. For general values of the parameters, the equation is not exactly integrable. In the present case, thanks to the presence of the imaginary term accompanying the first derivative, {analytic solutions can be obtained. Proceeding} by redefining
\bea
f_p (x) &=& g_p(x) e^{{\imath  \over 2} a_p x}
\eea
allows us to write Eq.~(\ref{eq_damped_cubed_oscillator}) as
\bea
g_p''(x) +  {\epsilon}^2_p g_p(x) - \gamma g_p^3(x) = 0,
\label{red_eq}
\eea
where we have defined
\bea
{\epsilon}^2_p = \left({a_p^2\over 4} + b_p \right).
\label{eq:47}
\eea
Equation (\ref{red_eq}) can be solved exactly in terms of Jacobi elliptic functions. 
We have taken advantage of the imaginary term in Eq.~(\ref{red_eq}) to reduce Eq.~(\ref{eq_damped_cubed_oscillator}) to a standard cubic {nonlinear Schr\"odinger} equation. This would have not been possible if the coefficient of the first derivative were any real number. 

{Since the model we are considering is classically unstable for $\lambda<0$, we shall focus here on the $\lambda>0$ case. The calculation of the quantization relations of the energy eigenvalues follows a similar approach to the nonrotating case; since it is slightly cumbersome and requires some {familiarity with} elliptic functions, the actual calculation will be relegated to the Appendix. Here, we summarize the main results. The solution to Eq.~(\ref{red_eq}) can be written as follows:
\bea
g_p(x) =  A\; \sn( qx +\delta, k^2),
\label{gpsn}
\eea
with the coefficients $A$, $\delta$, $q$ and $k^2$ to be determined by imposing the boundary and the normalization conditions, and by use of the first integral of Eq.~(\ref{red_eq}). As a result, $k^2$ (the elliptic modulus) and the eigenvalues $\epsilon_n^2$ are quantized according to the following equations (see the Appendix for details of the computation),
\bea
 {\lambda L\over 8(1-\beta^2) n^2}
&=&
 \Kappa(k^2_n) \left( {\Kappa(k^2_n)} - \Epsilon \left(k^2_n\right)\right),
\label{quant_m_n}\\
\epsilon_n^2 &=& {4n^2\over L^2} \Kappa^2(k_n^2) (1+k_n^2).
\label{quantz_eps_n}
\eea
with $0< n \in \mathbb{N}$, $\Kappa(k^2)$ and $\Epsilon(k^2)$ {defining the} \textit{complete elliptic integral of the first and second kinds}, respectively.}
{The spectrum of the rotating problem is encoded in the two equations (\ref{quantz_eps_n}) and (\ref{quant_m_n}). Thus, for any admissible value of the physical parameters, $L$, $\lambda$, $\beta$, and any positive integer $n$, Eq.(\ref{quant_m_n}) determines the value of the elliptic modulus $k_n^2$, which together with Eq.~(\ref{quantz_eps_n}) determines the eigenvalue $\epsilon_n$. Notice, that the angular velocity $\Omega$ ($\beta=\Omega R={\Omega L\over 2\pi}$) appears with even powers preserving the symmetry $\Omega \leftrightarrow -\Omega$.}

{The difference from the nonrotating case is rather nonobvious; rotation enters the frequencies $\omega_n$ via the coefficient $(1-\beta^2)$ in Eq.~(\ref{quant_m_n}) (see Eqs.~(\ref{Eq:42}), (\ref{Eq:43}) and (\ref{eq:47}) for the relation between $\omega_n$ and $\epsilon_n$). This is already nontrivial as the coupling constant is multiplied by the factor $1/(1-\beta^2)$, implying that the coupling constant can be effectively amplified if the rotation is large, i.e. when $\beta$ tends to unity. The second way that rotation enters the frequencies is via relation (\ref{quantz_eps_n}), which relates nonlinearly {to} the rotational velocity $\Omega$ {and} to the elliptic modulus $k^2_n$. That is, once the parameters ($\lambda$, $\beta$, $L$) and the quantum number $n$ are fixed, the elliptic modulus $k^2_n$ is determined according to Eq.~(\ref{quant_m_n}). Then, upon substitution into Eq.~(\ref{quantz_eps_n}), $k_n^2$ enters the frequencies.}

Before going into the computation of the vacuum energy, there are a number of relevant limits to check. The first is the noninteracting limit, $\lambda \rightarrow 0$. In this case, the first equation (\ref{quant_m_n}) gives
\bea
k_n^2 = 0, \mbox{$\forall n \in \mathbb{N}$.}
\eea
Substituting in Eq.~(\ref{quantz_eps_n}) we obtain
\bea
\omega_n \equiv \omega^{(0)}_n  =  {\pi n\over L}(1-\beta^2),
\eea
{which} coincides with the free rotating result (\ref{omega0n}). Naturally, taking the nonrotating limit, $\beta \to 0$, we also recover the nonrotating case.

Corrections to the above leading result are interesting in this case because of the way that {the} interaction strength and rotation intertwine. For small interaction strengths and zero rotation, the elliptic modulus can be obtained by expanding the right-hand side of Eq.~(\ref{quant_m_n}) for small argument. 
Ignoring terms of order $k_n^4$ or higher, we find
\bea
{ k^2_n} \approx {\lambda L\over \pi^2 n^2}.
\eea
Keeping only this first correction, we get for the frequencies
\bea
\omega^2_n 
&\approx& {{\pi^2}  n^2\over L^2} \left(1 + {3\over 2\pi^2}{\lambda L\over n^2} \right),
\eea
which recovers the result of Ref.~\cite{Bordag:2021}.

More interesting is the limit of small interaction strength with rotation on. In this case, relation (\ref{quant_m_n}) implies that the relevant expansion parameter is $\lambda/(1-\beta^2)$, which plays the role of an effective coupling. When rotation is small, $\beta \ll1$, then at leading order we return to the previous small-$\lambda$ small-$\beta$ case. When the ring is spinning fast, i.e. $1-\beta^2 \ll 1$, even for small $\lambda$ the quantity $\lambda/(1-\beta^2)$ can in principle be large. Thus we have two physically relevant limits. The first is for weak coupling and slow rotation, which yields the following expressions for the frequencies,
\bea
&& \omega^2_n {\approx} {{\pi^2} n^2\over L^2} \left(1 {+} {3\over 2\pi^2}{\lambda L\over n^2} \right)
\left(1 {-} \beta^2 \left( 3 {-} {2 n^2 \pi^2 \over 2 n^2 \pi^2 {+} 3\lambda L} \right)\right).\nonumber
\eea

The other is for weak coupling and fast rotation. In this case, the limit is trickier to take since in the calculation of the Casimir energy we need to sum over $n \in \mathbb{N}$. Since the sum extends to infinity, the quantity ${\lambda L/ 8/( 1-\beta^2)/n^2}$ is not necessarily large for any $n$  if ${\lambda L/( 1-\beta^2)}$ is large but finite. The limit of large but finite effective coupling, ${\lambda /( 1-\beta^2)}$, can be conveniently treated numerically, and we will do so in the next subsection. 

Finally, there is the case of ${\lambda /(1-\beta^2)} \to \infty$, i.e. infinitely strong coupling (which as explained can also be realized in the limit of small coupling and fast rotation); this corresponds in Eq.~(\ref{quant_m_n}) to the limit of $k_n^2 \sim 1$.  
Thus, from Eq.~(\ref{quant_m_n}) we get, for any $k_n^2$,
\bea
\Kappa (k^2_n)  = {1\over 2} \left(\Epsilon \left(k^2_n\right) \pm \sqrt{\Epsilon^2 \left(k^2_n\right) + {\lambda L^2\over 2 n^2 ( 1-\beta^2) }}\right).\nonumber
\eea 
Upon substituting in Eq.~(\ref{quantz_eps_n}) we get
\bea
\omega^2_n &=& {n^2\over L^2} \left(\Epsilon \left(k^2_n\right) \pm \sqrt{\Epsilon^2 \left(k^2_n\right) + 
{\lambda L^2\over 2 n^2 ( 1-\beta^2) }} \right)^2 \times\nonumber\\&& \times (1+k_n^2)(1-\beta^2)^2.
\eea
The leading term results from the $k_n^2 \to 1$ limit, yielding
\bea
\omega^2_n {\approx} {2 n^2 (1{-}\beta^2)^2\over L^2} \left(1 {+}\sqrt{1{+}{L^2 \lambda\over 2n^2 (1{-}\beta^2)}}\right)^2{+} \dots.
\eea
This regime {can also be} computed numerically.

\subsection{A numerical implementation of zeta regularization}

{In this section we shall compute the quantum vacuum energy $E_r$ given by the following (formal) expression
\bea
E_r = {1\over 2}\lim_{s\to 0} \sum_n{}^{'} \omega_n^{1-2s}.
\label{eq65}
\eea
The prime is a reminder that the sum is divergent and requires regularization.} 

{Before going into the specifics of the computation, it may be useful to explain the method we use to compute the vacuum energy in more general terms. The issue at hand is having to compute a summation of the form of Eq.~(\ref{eq65}) without any explicit knowledge of the eigenvalues. Knowing the eigenvalues would be enough to compute the above expression, were the summation convergent. However, as {is} common in quantum field theory, divergences are present and the above brute force approach cannot be used. The procedure described below is essentially a regularization procedure based on a numerical implementation of zeta function regularization.} 
\begin{figure}[t]
\includegraphics[width=1.0\columnwidth]{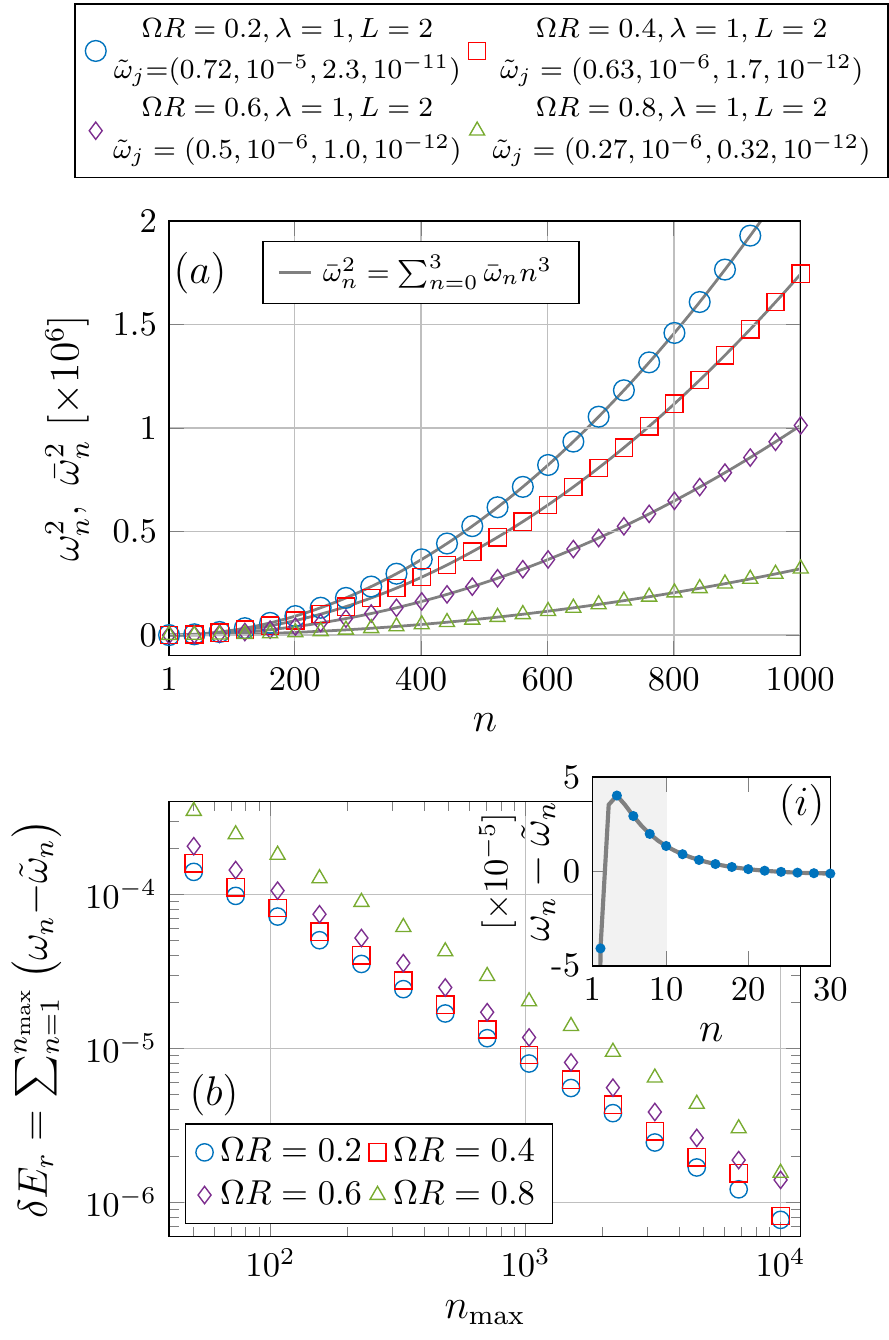}
{\caption{\label{fig:numeig}Numerical eigenvalue analysis. (a) Comparison of the polynomial fit (Eq.~\eqref{eqn:polyfit}) and the numerically computed eigenvalues (Eqs.~\eqref{quant_m_n} and \eqref{quantz_eps_n}) eigenvalues up to cubic order. The individual fitting coefficients are defined in the legend as $\tilde{\omega}_j=(\tilde{\omega}_0,\tilde{\omega}_1,\tilde{\omega}_2,\tilde{\omega}_3)$. (b) Correction $\delta E_r$ (Eq.~\eqref{eqn:der}) computed for each of the cases presented in (a), i.e. for fixed $\lambda=1$, $L=2$ with $\Omega R=0.2,0.4,0.6,0.8$. Inset: example of the eigenvalue difference $\omega_n-\tilde{\omega}_n$ for $\Omega R=0.2$.}}
\end{figure}
{Our approach consists in finding an approximate form for the eigenvalues, say, $\tilde\omega_n$, that can be used to compute the summation (\ref{eq65}) without relying on any approximation; this is possible {as long as the approximant $\tilde\omega_n$ captures the correct asymptotic behavior that causes the divergence}. In the present case, since we can find the eigenvalues numerically by solving Eqs.~(\ref{quant_m_n}) and (\ref{quantz_eps_n}) to any desired order and accuracy, we can proceed to find a suitable $\tilde\omega_n$ simply by fitting the eigenvalues. The form of the fitting function is unimportant (as we shall explain below), as long as the correct asymptotic behavior is captured by the fitting function. Assuming that this the case, we can write the above expression for $E_r$ as
\bea\label{eqn:caslong}
E_r {=} {1\over 2} \lim_{s\to 0}\sum_n \left(  \omega_n^{1-2s} {-} \tilde\omega_n^{1-2s}\right)
  {+} {1\over 2}\lim_{s\to 0}\sum_n \tilde\omega_n^{1-2s}. 
\eea
In the first term above the limit ${s \to 0}$ can be safely taken (by construction); we call this term
\bea\label{eqn:der}
\delta E_r ={1\over 2}\sum_n  \left( \omega_n - \tilde\omega_n\right).
\eea
{We note} that the above quantity can be computed numerically to any desired accuracy.}

{Thus, the Casimir energy {\eqref{eqn:caslong}} can be expressed as
\bea
E_r =  {1\over 2}\lim_{s\to 0}\sum_n \tilde\omega_n^{1-2s}+ \delta E_r.
\label{eqsplit}
\eea
There is no approximation at this stage: Eqs.~(\ref{eq65}) and (\ref{eqsplit}) are equivalent.} 

{Now, in order to get the \textit{exact} Casimir energy $E_r$ we are left with the computation of 
\bea
\tilde E_r =  {1\over 2}\lim_{s\to 0}\sum_n \tilde\omega_n^{1-2s},
\label{eqcas}
\eea
which, as detailed below, can be carried out analytically. Once we have $\tilde E_r$ and $\delta E_r$, these can be combined into an exact result for $E_r$.}

{There are still two issues to clarify. The first is about the renormalization of the Casimir energy: $\tilde E_r$ is still divergent and needs to be regularized and renormalized. In the present case the regularization, i.e. extracting the diverging behavior of the sum in $\tilde E_r$, is straightforward using zeta regularization. The renormalization also can be carried out at ease by subtracting from $\tilde E_r$ the asymptotic (i.e., calculated for $L \to \infty$) value of $\tilde E_r$ (practically, this is equivalent to {normalizing} the vacuum energy to zero in the absence of boundaries, a straightforward procedure in Casimir energy calculations).} 

{The other point to clarify regards the choice of the fitting function. First of all, one can easily understand that the choice of the fitting function is not unique, as two different fitting functions may result in differing $\delta E_r$. However, any such difference is irrelevant as it is compensated out in Eq.~(\ref{eqsplit}) due to the fact that we add and subtract the same quantity. Using two different fitting functions may result in an expedited numerical evaluation and in {an efficient} regularization of $\tilde E_r$.}

{Where uniqueness is required and two fitting functions cannot differ is in their asymptotic behavior. This is necessary to properly ``treat'' (i.e., renormalize) the vacuum energy. In order to find the asymptotic behavior one may proceed heuristically, essentially by trial and error, to find the best fitting function (this is straightforward to implement numerically in our case, or in situations where the eigenvalues can be computed numerically to any desired order and accuracy). In the present case, as discussed in the next section, we find the leading term in the frequencies $\omega_n^2$ behaves as $n^2$. A more general approach would be to invoke Weyl's theorem on the asymptotic behavior of the eigenvalues of a differential operator. In general dimensionality, given certain assumptions on the differential operator involved, the Weyl theorem provides {an intuitive} rationale for {obtaining} the correct asymptotic behavior as the eigenvalues behave as $n^{2/d}$ with $d$ being the dimensionality. To be precise, in its original formulation Weyl's law refers to the Laplace-Beltrami operator in $d\geq 2$, however, extensions of the theorem and to more general manifolds and operators, including in $d=1$, have been discussed in the literature \cite{Kirsten,Arendt,Bonder}.} 

{To recap, given two fitting function $\omega^{(fit)}_n$ and $\bar{\omega}^{(fit)}_n$ with the same asymptotic behavior, as {per} Eq.~(\ref{eqsplit}), the vacuum energy will not change: any change in the fitting function will result in different $\tilde E_r$ and $\delta E_r$, with the difference simply compensated by construction. Of course, as already stated, the arbitrariness in the choice of the fitting function can be used to achieve a speedier computation of $\delta E_r$ as well as a more manageable expression for $\tilde E_r$.}

{The above method will be concretely applied to our problem in the next section.}

\subsubsection{The Casimir energy}

In order to compute the vacuum energy we shall proceed as follows. First, we fix the values of the interaction strength $\lambda$, of the rotation parameter $\beta = {L \Omega\over 2\pi}$ and $L$ (the procedure is iterated over these parameters). Then, we solve Eq.~(\ref{quant_m_n})
for a sequence of $n \in \mathbb{N}$ up {to} some large value {$n_{\rm max}$}. These values of $k_n$ are then used to compute the frequencies $\omega_n$ up to {$n_{\rm max}$} according to formula (\ref{quantz_eps_n}).

With this set of frequencies $\Theta\left(n_*\right) = \left\{ \omega_1, \omega_2, \dots, \omega_{n_*}\right\}$, we fit the spectrum with a polynomial, 
\bea\label{eqn:polyfit}
\tilde \omega^2_n  =  \varpi_0 + \varpi_1 n + \varpi_2 n^2 + \varpi_3 n^3 + \dots + \varpi_{p_*} n^{p_*},
\eea 
with $2 \leq p_* \in \mathbb{N}$ (the fit is repeated for various values of $p_*$ with the value corresponding to the best fit selected. This is, in fact, a redundant step as it follows that the best-fit value is $p_*=2$). The next step consists in increasing the value $n_*$ and repeating the process until the coefficients $\varpi_k$ converge. {We find} that the best fit returns $\varpi_1 \approx 0$ and $\varpi_j \approx 0$ with $j\neq 2$, indicating that the spectrum is approximately of the form
\bea
\tilde \omega_n^2 \approx  \varpi_0 + \varpi_2 n^2,
\label{spectrumapprox}
\eea
where the ``$\approx$'' symbol should be understood in the sense of numerical approximation. 

As we explained in the preceding section, the form (\ref{spectrumapprox}) of the spectrum is anticipated from the expected Weyl-like asymptotic behavior of the eigenvalues, from which it follows $p_*=2$. As explained in the previous section, we should note that one may chose to proceed using a different fitting function with the same asymptotic behavior; this will not change the final result, but the following step should be modified accordingly. The choice {of} Eq.~(\ref{spectrumapprox}) is the easiest to handle and the most natural, {since the resulting summation will be of the form of a generalized Epstein-Hurwitz {zeta function (i.e., a type of zeta function associated with quadratic forms)} whose regularization can be carried out using a rearrangement by Chowla and Selberg, as explained below.}
\begin{figure}[t]
\centering
\includegraphics[width=1\columnwidth]{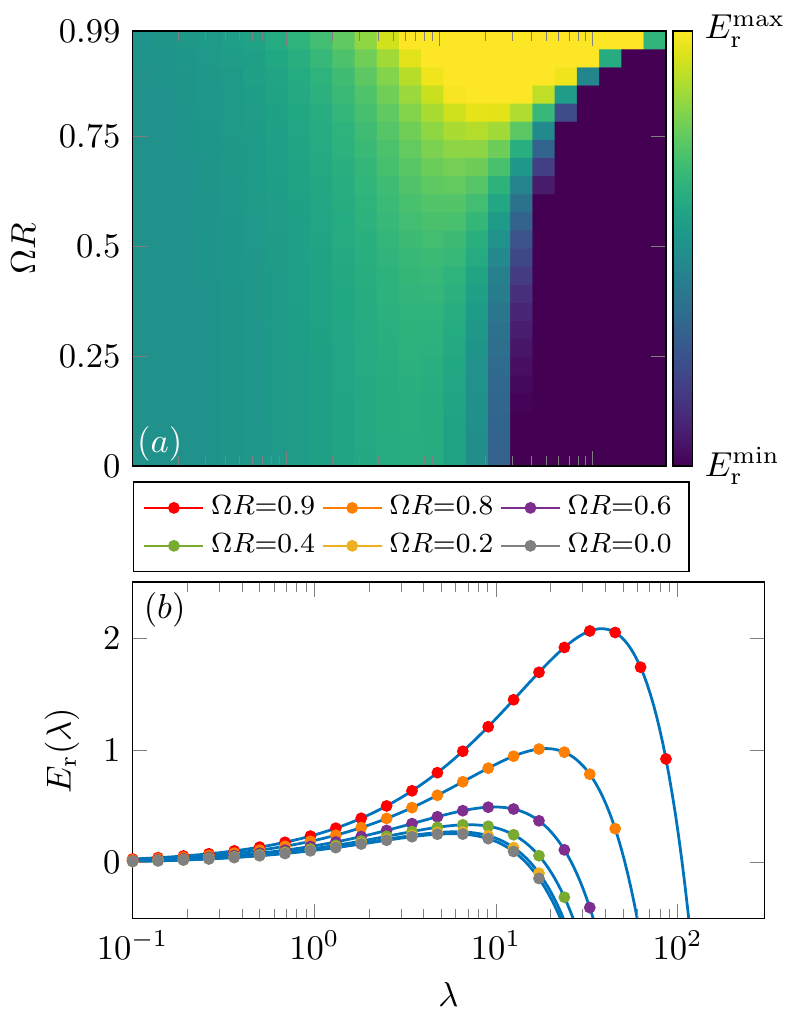}
\caption{\label{fig:2}Rotation-interaction heat map. (a) Casimir energy $E_r$ (Eq.~\eqref{eqn:er}) in the $(\Omega R,\lambda)$ parameter space. (b) $E_{r}(\lambda)$ for $\Omega R=0,0.2,0.4,0.6,0.8,0.9$. The size of the ring is $L=20$.}
\end{figure}
Once we have the eigenvalues $\omega_n$, these have to be summed according to Eq.~(\ref{eq65}) in order to obtain the Casimir energy of the rotating system.
\begin{figure*}[t]
\centering
\includegraphics[scale=0.8]{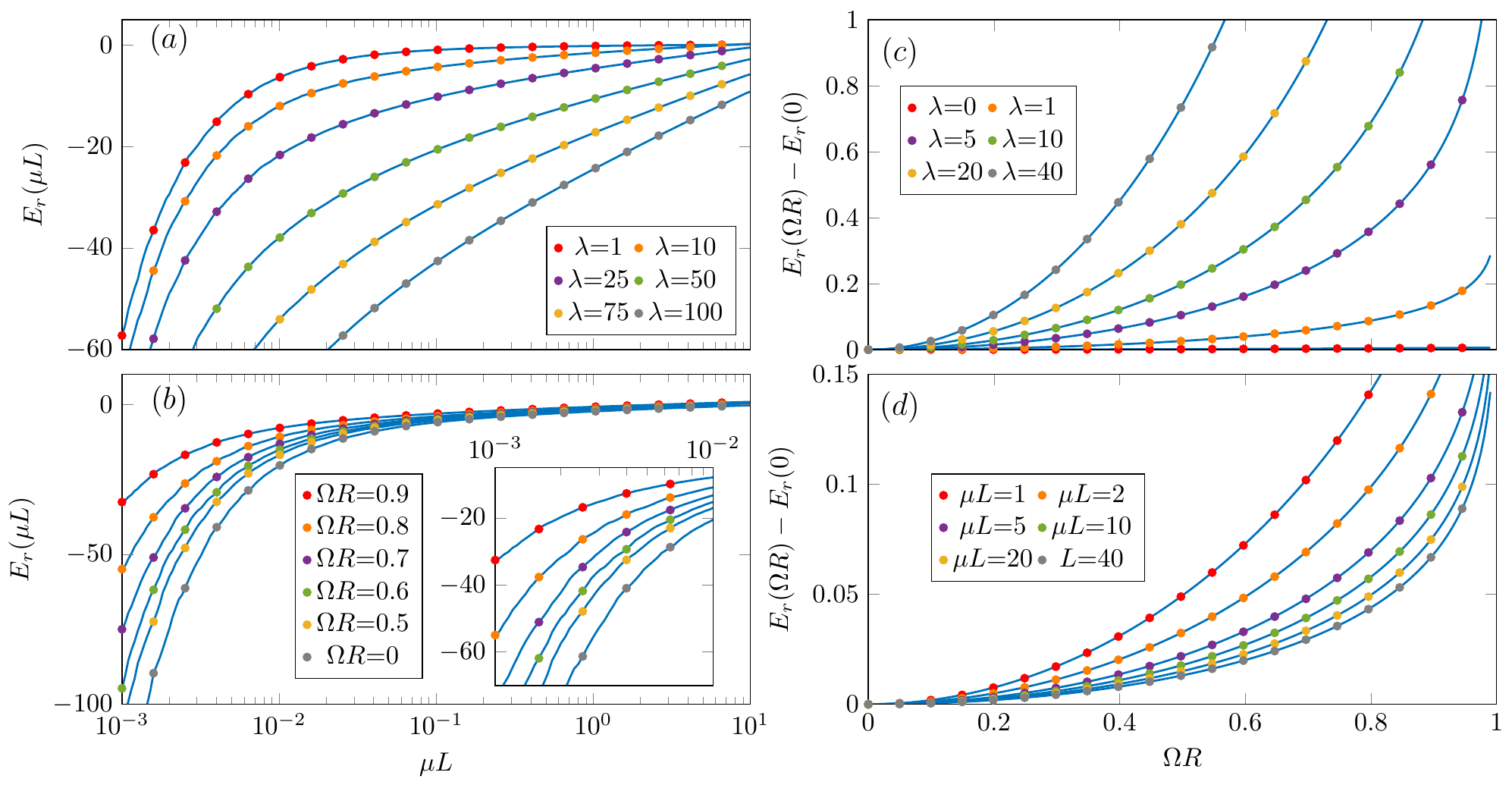}
\caption{\label{fig:grid}Casimir energies. (a) $E_r$ (Eq.~\eqref{eqn:er}) for varying $L$ for various fixed interaction strengths $\lambda=1,10,25,50,75,100$. The rotation parameter is $\Omega R=0.75$. (b) $E_r$ for fixed $\lambda=10$ for $\Omega R=0,0.5,0.6,0.7,0.8,0.9$. Inset: enlarged portion between $L=10^{-3}:10^{-2}$. (c), (d) $E_r$ as a function of the rotation parameter $\Omega R$ for a fixed ring length $L=20$ in (c) and a fixed interaction strength $\lambda=0.5$ in (d). {Here we have fixed $\mu=1$ throughout.}}
\label{figure3}
\end{figure*}
The computation of the Casimir energy is performed using the approximate formula {of} Eq.~(\ref{spectrumapprox}) leading to
\bea
\tilde E_r = \lim_{s\to 0} {\mu^{2s}\over 2}\varpi_2^{1/2-s} \sum_n \left({\varpi_0\over \varpi_2 } + n^2\right)^{1/2-s}.
\label{casimirexpl}
\eea
We have added the multiplicative term $\mu^{2s}$ in order to keep the dimensionality of the above expression to be that of energy \cite{Toms:2012}.

The expression (\ref{casimirexpl}) can be easily dealt with analytically when the ratio $\varpi_0 / \varpi_2$ is small, for instance in the noninteracting limit. In this case, a valid representation can be built by simply using the binomial expansion of the summand in Eq.~(\ref{casimirexpl}), yielding
\bea
\hskip -0.5 cm
\tilde E_r {=} -{\sqrt{ \varpi_2} \over 24}\left\{ 1 {+} 6{\varpi_0\over \varpi_2}\left[ 1 {-} \gamma_e {-} {1\over 2 s} 
{+} {1 \over 2} 
\log\left( \varpi_2 \over \mu^2 \right)\right]\right\} 
\eea
Notice that the noninteracting case corresponds to $\varpi_0 \to 0$, leading to
\bea
\tilde E_r &=& -{\sqrt{\varpi_2}\over 24},
\eea
implying that 
\bea
\lim_{\lambda \to 0} {\varpi_2} = {\pi^2 \over L^2}{\left(1 - \beta^2\right)^2}.
\eea
The above limit can (and has been) verified numerically.


When the ratio $\varpi_0 / \varpi_2$ is not small, we can use the following (Chowla-Selberg) representation for the sum above (see Ref.~\cite{Flachi:2008}):
\bea
\tilde E_r &=&
- {{\varpi_0}^{1/2} \over 4}\left\{1 - {1 \over 2} \sqrt{\varpi_0 \over \varpi_2} \left[ 1 - {\gamma_e} 
+ {1 \over s}  -  \log \left(\varpi_0 \over \mu^2\right) \right.\right.\nonumber\\
&&\left.
- {\psi(-1/2) } \right] \Big\}- {1 \over 2\pi}\sqrt{{\varpi_0}}\sigma\left( \sqrt{\varpi_0\over \varpi_2 }\right),\label{eqn:er}
\eea
where
\bea
\sigma(t) = \sum_{p=1}^\infty {1\over p}K_{1}\left( 2\pi p t \right).
\eea

For clarity we have left the diverging pieces $\propto 1/s$ in both expression for the vacuum energy. These terms are removed by subtracting {from} the nonrenormalized vacuum energy its counterpart for $L\to \infty$. We again stress here that using a different ansatz for the frequencies, for instance keeping a linear term, will change the expression (\ref{casimirexpl}) and, in turn, both representations for the summations have to be modified accordingly. However, as we have explained, this does not change the Casimir energy, since the difference is compensated away by the same change in $\delta E_r$.

In order to obtain the exact quantum vacuum energy, we need to add to $\tilde E_r$ the term $\delta E_r$ that can be calculated numerically,
\bea
\delta E_r = \lim_{s\to 0} {\mu^{2s}\over 2} \sum_n \left(\omega_n^{1-2s} - \tilde\omega_n^{1-2s} \right).
\eea
The above is nothing but the sum over the difference between the exact eigenvalues and the approximated ones. The advantage of using this formulation is that the above expression is regular in the ultraviolet and the limit $s \to 0$ can be taken without any further manipulation, as explained in the previous section. With the two pieces in hand the exact Casimir energy can be calculated as stipulated by Eq.~(\ref{eqsplit}).

{In Fig.~\ref{fig:numeig} we present an overview of our numerical methodology. Figure \ref{fig:numeig}(a) shows the eigenvalues of the polynomial fit (Eq.~\eqref{eqn:polyfit}) (gray solid) and the numerically computed ones (Eqs.~\eqref{quant_m_n} and \eqref{quantz_eps_n}) (open colored symbols), respectively. The comparison includes terms up to cubic order (viz. Eq.~\eqref{eqn:polyfit}) for fixed interaction strength and ring size, for various values of the rotation parameter $\Omega R$. Although the (odd) terms $\tilde{\omega}_1$ and $\tilde{\omega}_3$ have been included, they are many orders of magnitude smaller than $\tilde{\omega}_{0}$ and $\tilde{\omega}_{2}$, typically $\tilde{\omega}_1\sim 10^{-6}$ and $\tilde{\omega}_3\sim 10^{-12}$, supporting our choice of Eq.~\eqref{spectrumapprox}. We also checked the typical values of $\tilde{\omega}_{j>3}$, which are even smaller than the preceding terms. Then, Fig.~\ref{fig:numeig} (b) shows the quantity $\delta E_r$, Eq.~\eqref{eqn:der}. This quantity is calculated again for fixed $\lambda=1$, $L=2$ for various $\Omega R$, showing a monotonic decrease as $n_{\rm max}$ is increased from $n_{\rm max}=50:10^4$. Finally, the inset shows an example of the quantity $\omega_n-\tilde{\omega}_n$, displaying a prominent maximum located at $n\sim 5$. Due to the asymptotic nature of the regularization procedure, the behavior of this quantity is formally accurate for $n\gg 1$, hence we exclude the first $\sim 10\%$ of computed eigenvalues from our simulations (shaded gray), leading to the monotonic decrease of $\delta E_r$ toward zero for increasing $n_{\rm max}$. In the calculations that follow we set $n_{\rm max} = 1000$.}

\section{Results and discussion}

\begin{figure}
\includegraphics[width=0.95\columnwidth]{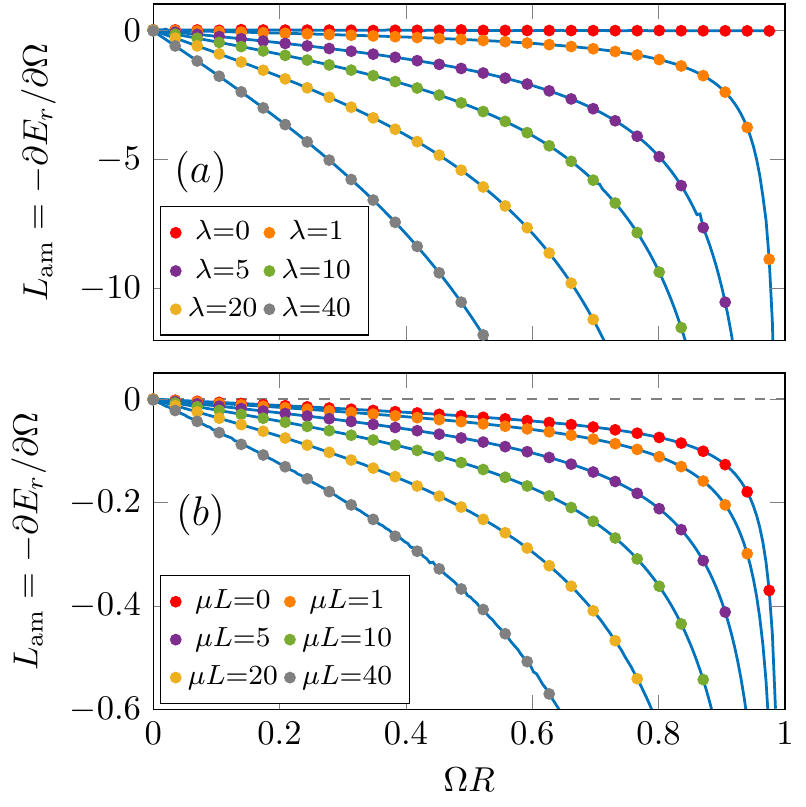}
{\caption{\label{fig:4} Angular momentum calculated per Eq.~\eqref{eqn:am}. (a) {$L_{\rm am}$} computed from the data in Fig.~\ref{fig:grid}(c) as a function of $\Omega R$ for fixed ring length $L=20$. (b) {$L_{\rm am}$} computed from the data in Fig.~\ref{fig:grid}(d) for fixed interaction strength $\lambda=0.5$.}}
\end{figure}

Figures~(\ref{fig:2}), (\ref{fig:grid}) and (\ref{fig:4}) illustrate the results. Figure \ref{fig:2}(a) summarizes in a rotation-interaction heat map how the Casimir energy $E_r$ depends on both the rotational parameter $\Omega R$ and the interaction strength $\lambda$. {A local maxima in the Casimir energy is present for $\Omega R\geq0$, as depicted in Fig.~\ref{fig:2} where Eq.~\eqref{eqn:er} is computed as a function of $\lambda$ for fixed values of $\Omega R$. We also found that increasing $L$, the size of the ring causes a proportional increase to the corresponding Casimir energy. Note that the existence of the local maxima depends on the size $L$ of the ring, appearing in the noninteracting $\Omega R=0$ limit for $L\gtrsim 5$.} 
In fact, the changes in the Casimir energy caused by both rotation and interaction are rather nontrivial, as we illustrate in Figs.~\ref{fig:grid}(a) and \ref{fig:grid}(b); Fig.~\ref{fig:grid}(a) shows the Casimir energy $E_r$ as a function of $L$ for various fixed and increasing interaction strengths $\lambda=1, \dots,100$ with rotation parameter set to $\Omega R=0.75$; Fig.~\ref{fig:grid}(b) shows $E_r$ as a function of $L$ for a fixed $\lambda=10$ and for increasing $\Omega R=0, \dots,0.9$. While the energy increases for rings of small sizes and decays to zero for large enough rings, from both cases it is clear that a departure from the typical $1/L$ dependence of the energy occurs as a consequence of both interaction and rotation (only rotation would not cause such a departure). Figures \ref{fig:grid}(c) and \ref{fig:grid}(d) illustrate the increase in the Casimir energy with respect to its nonrotating counterpart for different values of the interaction strength and for rings of different sizes. Finally, in Fig.~\ref{fig:4} we plot the angular momentum {$L_{\rm am}$} calculated according to Eq.~\eqref{eqn:am}: Fig.~\ref{fig:4}(a) shows {$L_{\rm am}$} computed from the data in Fig.~\ref{fig:grid}(c) as a function of $\Omega R$ for fixed ring length $L=20$ in (a) and from the data in Fig.~\ref{fig:grid}(d) for fixed interaction strength $\lambda=0.5$ in (b).

The Casimir effect, and more generally calculations of quantum vacuum energies for interacting field theories, have been considered for many years, starting with References \cite{Ford:1979b,Toms:1980a,Peterson:1982} (see also Ref.~\cite{Toms:2012}). Refs.~\cite{Maciolek:2007,Moazzemi:2007,Schmidt:2008,Flachi:2012,Flachi:2013,Flachi:2017,Chernodub:2017gwe,Chernodub:2018pmt,Valuyan:2018,Flachi:2017xat,Flachi:2021,Song:2021,Flachi:2022} give an incomplete list of more recent papers that have discussed different aspects of the Casimir effect and related physics in the presence of field interactions. Of particular interest to this work are Refs.~\cite{Flachi:2017,Bordag:2021} where the connection between the Casimir energy and elliptic functions has been pointed out and used, numerically in Ref.~\cite{Flachi:2017} and analytically in Ref.~\cite{Bordag:2021}, to compute the Casimir energy. Particularly worthy of notice are in fact the results of Ref.~\cite{Bordag:2021} that have highlighted clearly how to compute semianalytically (full analytic results can be obtained in specific regimes, but in general numerics cannot be avoided) and \textit{exactly} (without resorting to any approximation) the quantum vacuum energy dealing directly with the nonlinear problem. In this work, we have extended those results by adding rotation, a feature relevant to cold-atomic systems at least in the long wavelength regime where the perturbations around a Bose-Einstein condensate evolve according to a relativistic Klein-Gordon equation (see, for example, Refs~\cite{Kurita:2008fb,Barcelo:2010bq}). {Complementary to this, the Casimir effect has also been studied for superfluids in the presence of vorticity \cite{impens_2010,mendonca_2020}, as well as near  surfaces (Casimir-Polder interaction) \cite{pasquini_2004,harber_2005,moreno_2010,bender_2014}, at finite temperature \cite{obrecht_2007}, in the presence of disorder \cite{moreno_2010a} or with an impurity \cite{marino_2017}.} 

The other interesting aspect of the generalization we have considered here is that the mixing of interaction and rotation can give rise to a departure from the usual massless behavior of the Casimir energy for both the rotating and nonrotating noninteracting cases. Also, from Eq.~(\ref{quant_m_n}) one notices that rotation combines with the coupling constant in a way that even when the interaction strength is small, its effect can be amplified by fast rotation. Extending the present results to the nonrelativistic Gross-Pitaevski case can be done along the lines discussed here. More interesting would be generalizations to higher dimensions since the Coleman-Hohenberg-Mermin-Wagner theorem does not apply and phase transitions may occur dynamically without having to introduce any explicit breaking of rotational symmetry, {at least} in principle. In this case, not only will the Casimir effect experience a phase transition in correspondence to a critical value of the coupling constant and rotation at which symmetry breaking eventually occurs, but it may also become a proxy of critical quantities. This is of course a technically complicated problem as in more than one noncompact dimension the separability of the field equation becomes nontrivial. Obviously, the more interesting aspect to investigate is a closer connection to cold atoms and to the possibility of using these as a probe of quantum vacuum effects. 

\section*{ACKNOWLEDGEMENTS}
A.F.'s research was supported by the Japanese Society for the Promotion of Science Grant-in-Aid for Scientific Research (KAKENHI, Grant No. 21K03540). M.E.'s research was supported by the Australian Research Council Centre of Excellence in Future Low-Energy Electronics Technologies (Project No. CE170100039) and funded by the Australian Government, and by the Japan Society of Promotion of Science Grant-in-Aid for Scientific Research (KAKENHI Grant No. JP20K14376). One of us (A.F.) wishes to thank O. Corradini, G. Marmorini, and V. Vitagliano for earlier discussions and I. Moss for a comment regarding the use of a relativistic Klein-Gordon equation in Bose-Einstein condensates in the long wavelength approximation.

\appendix*

\section{\label{appA}Solutions and quantization of the energy eigenvalues}

{
In order to make the paper self-contained, in this appendix we present the details of the derivation of Eqs.~(\ref{quant_m_n}) and (\ref{quantz_eps_n}). The calculations below follow closely those of Refs.~\cite{Bordag:2021,Carr:2000::1,Carr:2000::2}, the difference being the inclusion of rotation. We should stress that although the calculations are similar to the nonrotating case (since we have reduced the equation to the form of Eq.~(\ref{red_eq})), the nonlinearity generated by the inclusion of rotation in the associated eigenvalue problem induces nontrivial changes in the frequencies, as will become clear below.} 

{
A general solution to Eq.~(\ref{red_eq}) with $\lambda>0$ can be written in terms of the Jacobi elliptic function $\sn(qx +\delta, k^2)$: 
\bea
g_p(x) =  A\; \sn( qx +\delta, k^2),
\label{gpsn1}
\eea
where $A$ is a normalization factor, $q$ a parameter, the elliptic modulus $k^2$, and the phase $\delta$; all these parameters are to be fixed by the boundary conditions, normalization and by a condition deriving from the first integral of the above equation (\ref{red_eq}). (For $\lambda<0$ a solution can be written in terms of $\cn(qx +\delta, k^2)$ (see Refs.~\cite{Bordag:2021,Carr:2000::2}), but we will not consider this case here.)} 

{Dirichlet boundary conditions at $x=0$ are satisfied by $\delta = 0$, leaving, at this stage, the other parameters undetermined; notice that it is possible to chose a different phase for $\delta$, however, this can be eliminated by a coordinate transformation. 
Imposing Dirichlet boundary conditions also at the other end $x=L$ gives 
\bea
&&f_p (x)\Big|_{x=0,L}=0, 
\eea
where $L = 2\pi R$, or explicitly, 
\bea
0 = \sn(q L, k^2).
\eea
The above relation implies a quantization for $q$,
\bea
q L = 2 \Kappa(k^2) n,
\label{quant_q}
\eea
with $0< n \in \mathbb{N}$ and $\Kappa(k^2)$ being the \textit{complete elliptic integral of the first kind} ($\Kappa(k^2)$ is quarter period of the Jacobi elliptic function $\sn$). Details on elliptic functions can be found in Ref.~\cite{nist}, while a discussion of the solutions to Eq.~(\ref{red_eq}) can be found in Ref.~\cite{Carr:2000::1} for the case of Dirichlet boundary conditions and $\lambda>0$ and in Ref.~\cite{Carr:2000::2} for the case of Dirichlet boundary conditions and $\lambda<0$. Below we go over the details for our case in a self-contained manner.} 

{To obtain the remaining parameters $k^2$ and $A$, we proceed as follows. After imposing the boundary conditions at $x= 0, L$ on the general solution, we can obtain the first integral associated with Eq.~(\ref{red_eq}), multiply Eq.~(\ref{red_eq}) by $g_p'(x)$, and integrate by parts,
\bea
\left(g_p'(x)\right)^{2}  + {\epsilon}^2_p g_p^2(x) - {1\over 2}\gamma g_p^4(x) = C.
\eea
with $C$ an undetermined constant. At this point we substitute the general solution (\ref{gpsn}) in the first integral above 
and equate like powers of $\sn$. This leads to the following relations:
\bea
{{\epsilon}^2_p} &=& (1+k^2) q^2, \label{eps2}\\
A^2 &=& 2 q^2 k^2 {1-\beta^2\over \lambda},\label{A2}\\
C &=& A^2 q^2. \label{C}
\eea
The last piece needed is the normalization of the solution, which allows us to fix the parameter $k^2$,
\bea
1 &=& \int_0^L \left|g_p(x)\right|^2 dx \nonumber\\
&=& {2 qL \over \gamma L}\left( {q L} -\mathcal{E} \left(\am(qL,k^2) , k^2 \right)\right),\nonumber
\eea
where $\mathcal{E}\left(x, k^2\right)$ is the \textit{Jacobi epsilon function} and $\am(x,m)$ is the \textit{Jacobi amplitude}. The Jacobi epsilon function can be defined as
\bea
\mathcal{E}\left(x, k^2\right) = \int_0^x \dn(z,k^2) dz;
\eea
the Jacobi amplitude is related to the Jacobi function by the following definition:
\bea
\am(x,k^2) = \int_0^x \dn(x',k^2) dx';
\eea
The function $\dn(x,k^2)$ is the Jacobi delta amplitude, related to the $\sn$ function by
\bea
\dn(x,k^2) = \sqrt{1-k^4 \sin^2\left(\am(x, k^2)\right)}.
\eea 
The Jacobi amplitude satisfies the following relation,
\bea
\am(2 n \Kappa(k^2), k^2) = n \pi
\eea
which allows the normalization condition to be rewritten as
\bea
1 &=&
 {4 \Kappa(k^2) n \over \gamma L}\left( 
{2 \Kappa(k^2) n} 
-\mathcal{E} \left(\am(2 \Kappa(k^2) n,k^2) , k^2 \right)
\right).\nonumber
\label{norm}
\eea
Using the identity 
\bea
\mathcal{E} \left(n\pi, k^2\right) = 2 n \Epsilon \left(k^2\right),
\eea
with $\Epsilon \left(k^2\right)$ being the \textit{complete elliptic integral of the second kind}, together with the periodicity of the Jacobi $\sn$, Eq.~(\ref{quant_q}), we arrive at the following relation
\bea
 { \lambda L\over 8(1-\beta^2) n^2}
&=&
 \Kappa(k^2_n) \left( {\Kappa(k^2_n)} - \Epsilon \left(k^2_n\right)\right),
\label{quant_m_n_2}
\eea
Notice that the generic index $p$ is now quantized, $p \to n$ and so is the elliptic parameter $k^2$. Substituting $k^2 \to k_n^2$ and $q \to q_n$, as {per} (\ref{quant_q}), we can also get the quantization relation for the eigenvalues $\epsilon_n$:
\bea
\epsilon_n^2 = {4n^2\over L^2} \Kappa^2(k^2) (1+k^2).
\label{quantz_eps_n_2}
\eea
}

\end{document}